\documentclass[]{zakoproc}
\usepackage{graphicx}
\begin{document}

\title{Recent achievements on KMHD and applications in the ICM}
\author{Diego Falceta-Gon\c calves}
\institute{Escola de Artes, 
Ci\^encias e Humanidades, 
Universidade de S\~ao Paulo, Rua Arlindo Bettio 1000, CEP 03828-000,
S\~ao Paulo, Brazil}
\markboth{D. Falceta-Gon\c calves}{Recent achievements on KMHD \ldots}

\maketitle

\begin{abstract}
Collisionless plasmas are ubiquitous in the Universe. In general full
kinetic equations are needed for a correct description of the physics in such
environments. Fluid approximations are, at some special cases, valid. In
this work we present the basic equations under such an approximation and the implementation of them in a magnetohydrodynamical code. The numerical simulations are applied for intracluster medium physical properties, i.e. low density but high temperature plasma. We show that, under specific conditions firehose and mirror instabilities may be present. The side effects are the conversion of thermal to magnetic/kinetic energies and the generation of small scale structures ($\sim 1$kpc). We study the changes in the statistics of Faraday rotation measurement, and of the synchrotron polarization, when compared to a standard MHD simulation for the ICM. We conclude that KMHD numerical simulations are mandatory in order to fully understand the statistics of the ICM emission. 
\end{abstract}

\section{Introduction}

The physical properties of the interstellar medium, such as mean magnetic field ($B \sim \mu$G), temperature ($T \sim 20 - 10^4$K) and density ($10-10^4$cm$^{-3}$), result in mean-free paths ($\lambda_{\rm mfp} \sim 10$AU) much smaller than typical dimensions of the system ($L \sim 100$pc). It means that the ISM plasma may be well described under fluid approximation. Also, since the collision frequency of ions is typically much larger than their cyclotron frequency ($\nu_{\rm ii} \gg \Omega_{\rm i}$), we may also assume the momentum distribution to be isotropic (Maxwellian). 

In the intracluster medium, on the other hand, with ($B \sim 0.1 \mu$G), temperature ($T \sim 10^8$K) and density ($10^{-2}$cm$^{-3}$), the collision frequency of ions is typically much smaller than their cyclotron frequency ($\nu_{\rm ii} \ll \Omega_{\rm i}$), though its mean-free path ($\lambda_{\rm mfp} \sim 1$kpc) is still smaller than the typical dimensions of the system ($L \sim 0.1 - 1$Mpc). Therefore, even though the ICM being properly described under fluid approximation, its momentum distribution will no longer be isotropic. These plasmas are often called gyrotropic and the momentum distribution of particles is usually described by a bi-Maxwellian function. One for the motions parallel to the magnetic field, and another for the perpendicular. In other words, a gyrotropic plasma presents two temperatures ($T_\parallel$ and $T_\perp$), with respect to the local magnetic field.

In this case, some dynamical properties of the plasma may change. The evolution of the turbulent cascade will be different, when compared to that of a standard MHD turbulence, for instance. Also, other kinetic effects, such as instabilities may take place. If these effects are dominant, the understanding of the physical properties of the ICM may be wrong. This issue is briefly discussed in this work.

\section{The KMHD model}

In the fluid approximation, a gyrotropic plasma can be described by the CGL-magnetohydrodynamic (Chew
{\em et al.} 1956) equations expressed as follows (see also Kowal, Falceta-Gon\c calves \& Lazarian 2011):

\begin{eqnarray}
 \frac{\partial \rho}{\partial t} + \nabla \cdot \left( \rho \bf{v} \right) & = & 0, \label{eq:mass} \\
 \frac{\partial \rho \bf{v}}{\partial t} + \nabla \cdot \left[ \rho \bf{v} \bf{v} + \left( \mathsf{P} + \frac{B^2}{8 \pi} \right) I - \frac{1}{4 \pi} \bf{B} \bf{B} \right] & = & \bf{f}, \label{eq:momentum} \\
 \frac{\partial \bf{B}}{\partial t} - \nabla \times \left( \bf{v} \times \bf{B} \right) & = & 0, \label{eq:magnetic_flux}
\end{eqnarray}

\noindent
where $\rho$ and $\bf{v}$ are the plasma density and velocity, respectively,
$\bf{B}$ is the magnetic field, $\mathsf{P} = p_\perp \hat{I} + (p_\parallel -
p_\perp) \hat{b} \hat{b}$ is the pressure tensor, $\hat{b}=\bf{B}/|\bf{B}|$ is
the unit vector along the magnetic field, $p_\parallel$ and $p_\perp$ are the
pressure components parallel and perpendicular to $\hat{b}$, respectively, and
$\bf{f}$ represents the turbulent forcing term. Further details on the code may be obtained in Falceta-Gon\c calves et al. (2010a,b), Falceta-Gon\c calves, Lazarian \& Houde (2010).

The above set of equations is closed by the determination of the parallel and
perpendicular pressures. Here, we make use of the double-polytropic equations, in conservative form, as:

\begin{eqnarray}
 \frac{\partial \left(p_\perp B^{1 - \gamma_\perp}\right)}{\partial t} + \nabla \cdot \left( p_\perp B^{1 - \gamma_\perp} \bf{v} \right) & = & 0, \\
 \frac{\partial \left( p_\parallel
\left( B / \rho \right)^{\gamma_\parallel-1} \right)}{\partial t} + \nabla \cdot \left( p_\parallel
\left( B / \rho \right)^{\gamma_\parallel-1} \bf{v} \right) & = & 0,
\end{eqnarray}

where $\gamma_\perp$ and $\gamma_\parallel$ are the polytropic exponents for the
perpendicular and parallel pressures, respectively.  

In this work we focus on the double-isothermal case, i.e. the pressure ratio $c_\perp / c_\parallel$ is constant (Hasegawa 1969). In this way we can observe the role of the instabilities during the whole simulation. Typically, the firehose and mirror instabilities grow fast, at small scales mainly (see Figure 1), and the pressure anisotropy tends to be reduced very quickly. Several natural processes, such as scattering of cosmic rays, or anisotropic thermal conduction, may lead to the unstable regimes, but we avoided dealing with these complications and focus only on the unstable regime. 

In Figure 1 we show the turbulent velocity spectrum of the simulated plasmas at different unstable regimes, firehose and mirror. Compared to the standard MHD case, it is clear the increase in the total power at small scales due to the instabilities. The mirror unstable case is more prominent since it acts at the magnetosonic modes (compressible), while firehose at the Alfvenic mode.

\begin{figure}[ht]
 \includegraphics[width=1.0\textwidth]{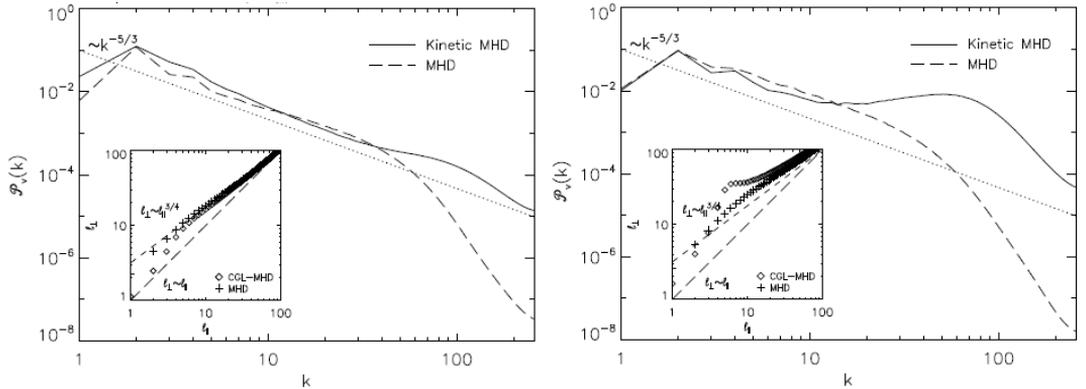}
 \caption{Velocity power spectrum obtained for the KMHD firehose-unstable model (left), with $c_\perp / c_\parallel = 0.5$, and for the KMHD mirror-unstable case (right) with $c_\perp / c_\parallel = 2.0$. MHD case is shown as dashed line for comparison (extracted from Kowal, Falceta-Gon\c calves \& Lazarian 2011).}
 \label{a}
\end{figure}

\section{The KMHD effects in the ICM}

Among the main observational methods to infer the ICM magnetic fields are the synchrotron polarization and the Faraday rotation measurements (e.g. Giovannini, Tordi \& Ferreti 1999, Ensslin \& Vogt 2003, Govoni \& Feretti 2004, Guidetti et al. 2010 and others).

The intrinsic large polarization degree of the synchrotron emission makes it very useful in studying the ICM magnetic fields. Polarization maps reveal the field line structure in the plane of sky. Complex, non-uniform, distributions reveal a kinetic dominated plasma, where $B_{\rm sky}^2 < \rho \delta v^2 \sim \delta B^2$. Very uniform polarization vectors, on the other hand, reveal strong magnetic fields with $B_{\rm sky}^2 >> \rho \delta v^2$ (Falceta-Gon\c calves, Lazarian and Kowal 2008). Unfortunately, this technique does not tell much about the magnetic field component parallel to the line of sight. In this particular case, the Faraday rotation effect may be useful.

\begin{figure}[ht]
 \includegraphics[width=0.9\textwidth]{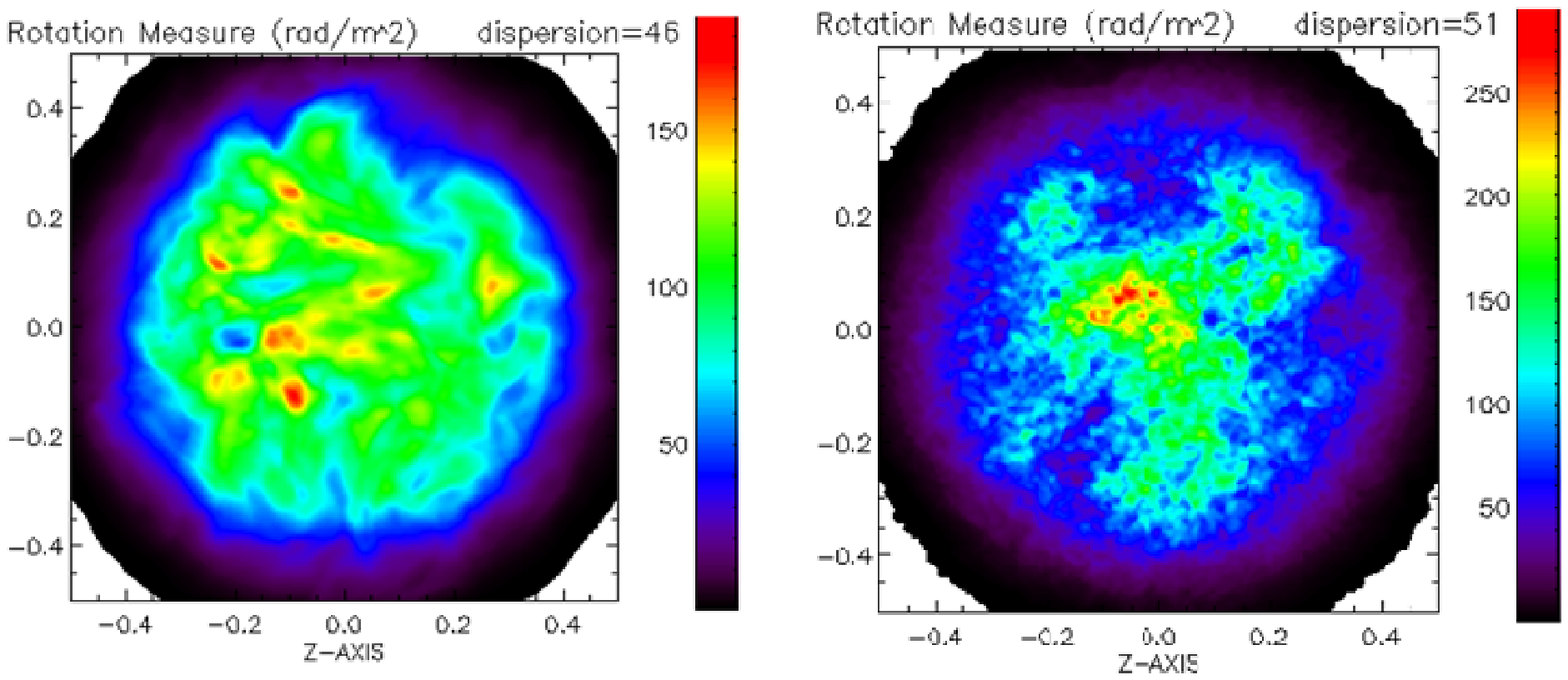} \\
 \includegraphics[width=0.9\textwidth]{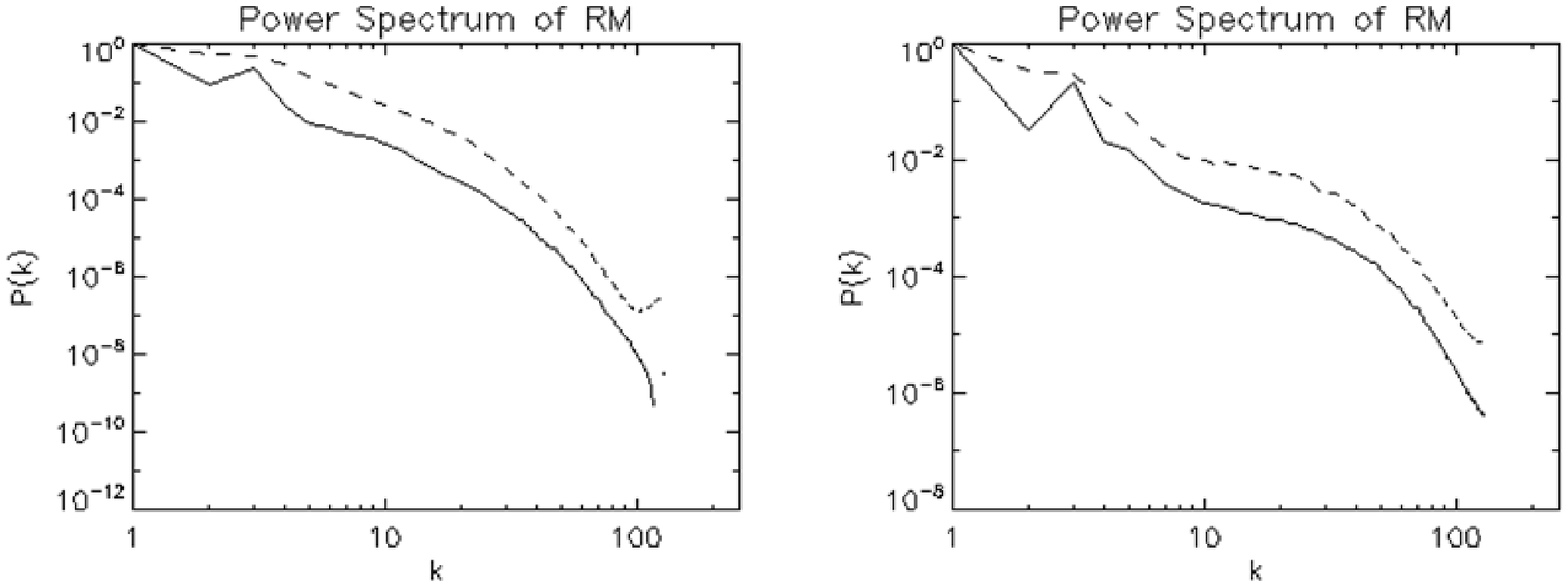}
 \caption{Up: Rotation measure calculated for the MHD model (left), with $B_0 = 0.1 \mu$G, $L = 1$Mpc and $n_0 = 0.01$cm$^{-3}$, and for the KMHD mirror-unstable case (right) with $c_\perp / c_\parallel = 2.0$. Bottom: power spectrum of the spatial distribution of the rotation measure calculated for the MHD model (left), with $B_0 = 0.1 \mu$G, $L = 1$Mpc and $n_0 = 0.01$cm$^{-3}$, and for the KMHD mirror-unstable case (right) with $c_\perp / c_\parallel = 2.0$.}
 \label{b}
\end{figure}

Most of the radio sources in clusters of galaxies are embedded in magnetized, hot and low density plasmas that account for the frequency dependent refraction index, which is responsible for the Faraday rotation. The polarization angle of the observed radiation depends on the frequency as follows:

\begin{eqnarray}
\phi_{\rm obs} (\lambda) = \phi_{\rm 0} + \lambda ^2 \times RM {\rm , where} \\
RM \simeq 812 \int^{L(kpc)}_{0}{n_e(cm^{-3}) B_{\parallel}(\mu G) dl} \ \ \ ({\rm rad \ m}^{-2}).
\end{eqnarray}

Once the distribution of matter ($n_e$) is known - from X-rays maps, for example - it is possible to extract information about $B_{\parallel}$. Combined to the polarization maps, the rotation measure (RM) is a good tracer of the magnetic field in the ICM, and its coupling with the local plasma.

Galactic motions within the cluster medium is expected to occur at scales of $\sim 1$Mpc, with dispersion of velocities of order of $1000$ km s$^{-1}$. This kinetic energy density is about an order of magnitude larger than the energy density of the average uniform magnetic field component. Therefore, the turbulence in the ICM tends to be responsible for changes in the structure of the field lines as well.

The statistical study of the RM observed in clusters of galaxies has also been used in order to understand its structure and coupling with the turbulent medium (Ensslin \& Vogt 2003). Usually, the observed RM maps are compared to cosmological simulations that are, in essence, MHD. In a more realistic case, under KMHD regime, the structure of the magnetic field and its role on the propagation of turbulent modes may change. This may be obtained from the observed RM maps. In Figure 2 we show the polarization maps obtained for the two regimes (left:MHD, right:KMHD - mirror unstable), considering $L = 100$kpc, $B_{0}=0.1 \mu$G and $<n_e> = 0.005$. The difference in structure is clear. The KMHD mirror-unstable case shows a power increase at small scales, as seen in the power spectra (Figure 2, bottom panels). Also, the rotation measure in Figure 2 is larger for the KMHD case. That occurs because the mean magnetic field is in the plane of sky. The stronger perturbations in the KMHD case bend the field lines towards the LOS direction, increasing the RM. This difference will be reduced as the angle between the mean field and the plane of sky increases.

\begin{figure}[ht]
 \includegraphics[width=1.0\textwidth]{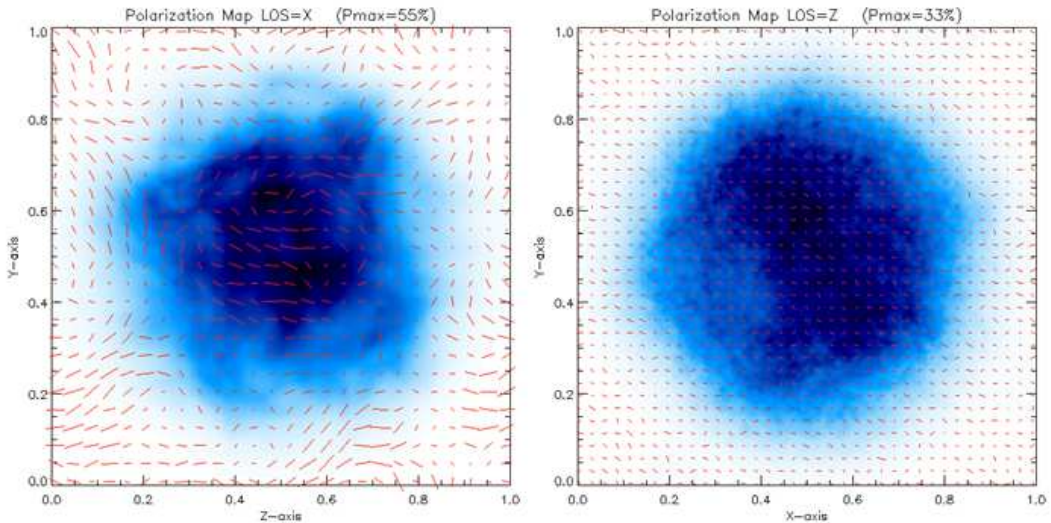}
 \caption{Polarization maps of the synchrotron emission calculated for the MHD model (left), with $B_0 = 0.1 \mu$G, $L = 1$Mpc and $n_0 = 0.01$cm$^{-3}$, and for the KMHD firehose-unstable case (right) with $c_\perp / c_\parallel = 0.5$.}
 \label{d}
\end{figure}

The polarization maps for both cases are shown in Figure 3. The maximum polarization degree obtained for the MHD case is 55\%. The intrinsic synchrotron polarization, around 70\%, is reduced due to the turbulent structure of the ICM. The increase of the random component of the magnetic field increases the dispersion of the polarization angle ($\phi$) and reduces the polarization degree. For the KMHD case we see two different situations. For the mirror-unstable regime the instabilities produce compressions that are not in the Alfvenic modes, so they do not increase the random component of the field. In the firehose-unstable regime, on contrary, the instability operates at the Alfvenic modes and result in an increase in the random component of the field. As a result, as shown in Figure 3 (right), the polarization degree decreases to $\sim 33\%$.

\section{Conclusions}

Main results are:
\begin{itemize}
\item[$\bullet$]{
kinetic effects may change the dynamics of the ICM plasma;}
\item[$\bullet$]{
depending on the regime, fast increase of magnetic field may take place;}
\item[$\bullet$]{
as well as acceleration (fluid motions), both at small scales ($\sim 1$kpc);}
\item[$\bullet$]{
the KMHD instabilities result in modifications of statistics of RM and synchrotron polarization.}
\end{itemize}

In the double adiabatic case, the differences between the standard MHD approach and the one presented in this work will, obviously, be less prominent since one of the effects of the instabilities is to decrease pressure anisotropy. However, we believe that most of the conclusions, mostly regarding fast B-growth and modifications in the statistics of RM and synchrotron polarization remain. This will be tested in the near future with proper simulations.

\acknowledgements{The author thanks the Brazilian agency FAPESP (no. 2009/10102-0) for financial support.}

\end{document}